\documentstyle[aps,prb,floats,epsf,epsfig]{revtex}

\begin{document}
\input epsf
\draft

\renewcommand{\floatpagefraction}{1.00}
\renewcommand{\topfraction}{1.00}
\renewcommand{\textfraction}{0.00}
\renewcommand{\bottomfraction}{1.00}

\twocolumn[\hsize\textwidth\columnwidth\hsize\csname@twocolumnfalse%
\endcsname
\tightenlines
\draft

\title {{\rm\small\hfill to appear in Phys. Rev. B}\\
Stability of sub-surface oxygen at Rh(111)}

\author{M.Veronica Ganduglia-Pirovano \cite{humboldt}, Karsten Reuter, and Matthias Scheffler}

\address{Fritz-Haber-Institut der Max-Planck-Gesellschaft, Faradayweg
4-6, D-14195 Berlin, Germany}

\date{\today}
\maketitle

\begin{abstract}
Using density-functional theory (DFT) we investigate the incorporation of
oxygen directly below the Rh(111) surface. We show that oxygen incorporation
will only commence after {\it nearly} completion of a dense O adlayer 
($\theta_{\rm tot}\approx 1.0$ monolayer) with O in the fcc on-surface sites.
The experimentally suggested octahedral sub-surface site occupancy, inducing
a site-switch of the on-surface species from fcc to hcp sites, is indeed 
found to be a rather low energy structure. Our results indicate that at even
higher coverages oxygen incorporation is followed by oxygen agglomeration in 
two-dimensional sub-surface islands directly below the first metal layer. 
Inside these islands, the metastable hcp/octahedral (on-surface/sub-surface) 
site combination will undergo a barrierless displacement, introducing a 
stacking fault of the first metal layer with respect to the underlying
substrate and leading to a stable fcc/tetrahedral site occupation. We suggest 
that these elementary steps, namely, oxygen incorporation, aggregation into
sub-surface islands and destabilization of the metal surface may be more 
general and precede the formation of a surface oxide at close-packed late 
transition metal surfaces.
\end{abstract}
]

\narrowtext
\vskip2pc

\section{Introduction}
Exposing a metal surface to oxygen can result in simple adsorbate covered
surfaces, O sub-surface penetration or oxide formation, depending markedly on
the oxygen partial pressure, substrate temperature, the crystallographic
orientation of the surface, and the time of exposure. As the chemical activity
of the metal surface in these different states can vary significantly, knowledge
of the interaction of oxygen with metal surfaces is critical for understanding
technologically important processes like e.g. oxidation catalysis. The formation
of surface oxides on metal surfaces can in this respect both be beneficial, as
well as detrimental. In particular, for applications such as the CO oxidation on
transition metal based catalysts, oxide formation at the catalyst's surface in
the reactive environment was primarily viewed as leading to an inactive surface
oxide poisoning the catalytic reaction, as e.g. in the case of Rh
surfaces,\cite{oh83,kellogg85,peden88} while recent experiments showed that the
high catalytic activity of the Ru(0001) surface with respect to the CO oxidation
relates in fact to the existence of RuO${}_2$(110) oxide patches that form under
oxidizing conditions. \cite{boettcher97,over00,kim01} Thus, apparently,
unreactive surface oxides are more likely to form on Rh than on Ru surfaces.
These findings call for an atomistic understanding of the oxidation of these
transition-metal surfaces and of the structure of their oxide surfaces.

There is general agreement that the mechanistic steps leading to oxide formation
involve dissociative oxygen chemisorption on the metal surface, lattice
penetration of atomic oxygen, and the crystallization and growth of the 
stoichiometric oxide phases. The sequence of these events may be complex, and
rather than successively, they may occur simultaneously, depending on the 
temperature and pressure. The first of these steps, i.e. the chemisorption
of oxygen on Rh single crystal surfaces, has been the subject of intensive
experimental and theoretical research,
\cite{comelli98,loffreda98,pirovano99,pirovano01} but only few experimental
studies have addressed the ensuing formation of sub-surface oxygen,
\cite{thiel79,rebholz92,peterlinz95,gibson95,janssen96,wider99,gibson99}.
In one of these studies it could recently be
shown \cite{wider99} that exposure of the Rh(111) surface to O$_2$ at
moderatly elevated temperatures ($\sim 470$ K) leads to the formation of
sub-surface oxygen species, and there is evidence that at least $\sim 0.9$
monolayer (ML) of O is adsorbed on the surface before sub-surface sites are
occupied. Interestingly, the analysis of the X-ray photoelectron diffraction
(XPD) data of the corresponding study suggested that the sub-surface species
($\sim 0.1$ ML) occupies octahedral sites between the first and second Rh 
outermost layers, which lie just underneath the fcc on-surface adsorption
sites, while the neighboring on-surface oxygen has {\it switched} from 
its normal fcc to the hcp sites.

On theory side, {\em ab initio} studies of O sub-surface species on 
transition metal surfaces are even more scarce 
\cite{kiejna00,reuter02a,reuter02b} and as yet lacking for Rh surfaces.
The present theoretical study therefore specifically examines
the incorporation of O into the close-packed Rh(111) surface using 
density-functional theory. Extending preceding work focusing on ordered
oxygen phases {\em on} this surface \cite{pirovano99,pirovano01} we now 
discuss the stability of oxygen {\em below} the surface, and address 
questions concerning the minimum oxygen coverage at the onset of oxygen 
penetration (section III.A), the stability of various available sub-surface 
sites (section III.B), as well as the effect of a continued oxygen 
incorporation with increasing coverage up to 2.0 ML (section III.C). The
present results for sub-surface oxygen on Rh(111) are discussed and compared 
with corresponding results for the Ru(0001) surface \cite{reuter02a,reuter02b}
as a first step towards the understanding of the elementary steps governing
the onset of surface oxide formation. Questions remain as to the nature of
the distinct chemical activity of oxidized Ru(0001) and Rh(111) surfaces.

\section{Calculational details}\label{sectionII}

All calculations have been performed using
density-functional theory (DFT) and the generalized
gradient approximation (GGA) of Perdew {\em et al.}
\cite{perdew92} for the exchange-correlation functional
as implemented in the full-potential linear augmented
plane wave method (FP-LAPW) \cite{blaha99,kohler96,petersen00}.
The Rh(111) surface is modeled in the supercell approach,
employing a 7-layer (111) Rh slab with a vacuum region
corresponding to 6 interlayer spacings ($\approx 13$\,{\AA}).
Oxygen atoms are adsorbed on both sides of the slab. Their
positions as well as those of all Rh atoms in the two
outermost substrate layers are allowed to relax while
the central three layers of the slab are fixed in their
calculated bulk positions. In a preceding publication
we have detailed the geometrical properties and stability
of ordered adlayers of O on Rh(111) \cite{pirovano99}.
Exactly the same calculational setup is used here, such
that the results of our prior study can be used for
comparison. Further technical details of the calculations
can correspondingly be found in Ref. \onlinecite{pirovano99}.

We address the stability of O/Rh(111) structures with respect
to adsorption of $\rm O_2$ by calculating the average
adsorption energy per O adatom,

\begin{equation}
\label{eq1}
E_{\rm ad}(\theta_{\rm tot}) = E_{\rm b}(\theta_{\rm tot}) - D/2\ ,
\end{equation}

\noindent
where $D$ is the dissociation energy of the $\rm O_2$ molecule
and $E_{\rm b}(\theta_{\rm tot})$ the average binding energy
per oxygen atom as a function of the total coverage (i.e., on- + sub-surface)
$\theta_{\rm tot}$. In turn, $E_{\rm b}(\theta_{\rm tot})$ is defined as

\begin{eqnarray}
\label{eq2}
E_{\rm b}(\theta_{\rm tot}) &=& - \frac{1}{N_{\rm tot}}
[\; E_{\rm O/Rh(111)}^{\rm slab}(\theta_{\rm tot}) - \\ \nonumber
&&(E_{\rm Rh(111)}^{\rm slab} + N_{\rm tot} E_{\rm O}^{\rm atom}) \;]\ ,
\end{eqnarray}

\noindent
where $N_{\rm tot}$ is the total number of oxygen atoms in the unit-cell, and
$E_{\rm O/Rh(111)}^{\rm slab}$, $E_{\rm Rh(111)}^{\rm slab}$, and $E_{\rm
O}^{\rm atom}$ are the total energies of the O/Rh(111) adsorbate system, of the
clean Rh(111) surface, and of the free oxygen atom, respectively. A positive
value of the average adsorption energy indicates that the dissociative
adsorption of O$_2$ is exothermic. That is, the binding energy per O adatom on
Rh(111) is larger than that which the O atoms have in $\rm O_2$(gas), i.e.,
$D/2$. The total energies of the adsorbate system and clean surface are
calculated using the same supercell. Details on the calculations of the isolated
O atom and the free $\rm O_2$ molecule are given in Ref.
\onlinecite{pirovano99}.

To test the accuracy of the calculated binding energies, $E_{\rm b}(\theta_{\rm
tot})$, on numerical approximations due to the finite FP-LAPW basis set and the
finite slab and vacuum thicknesses in the supercell approach, selected
calculations were repeated with higher accuracy. The self-consistent
calculations of $E_{\rm b}$ were routinely conducted for a 7-layer slab, a 16 Ry
plane-wave cutoff in the interstitial region, and a $(12\times 12\times 1)$
Monkhorst-Pack grid for the $(1\times 1)$ unit-cell with 19 {\bf k}-points in
the irreducible wedge \cite{pirovano99}. Changing to denser {\bf k}-meshes up to
a $(18\times 18\times 1)$ grid with 37 {\bf k}-points in the irreducible wedge
resulted in negligible variations of $E_{\rm b}$ within $\sim 10$ meV/O atom.
Similarly, extending the Rh(111) slab from 7 to 9 and 11 layers led only to
changes in the calculated binding energies up to $\sim 10$ meV/O atom. The only
really notable effect on the computed $E_{\rm b}$ is caused by variations of the
finite plane-wave cutoff in the interstitial region: Increasing this cutoff from
16 Ry to 24 Ry decreases the binding energies by $\approx 100$ meV/O atom. Yet,
this decrease can be largely attributed to an improved description of both the
free O atom and the chemisorbed species and thus affects all structures that
contain the same amount of O alike. Consequently, the calculation of
relative stabilities of different phases, i.e. the difference between two
binding energies, has a significantly smaller error than this variation in the
absolute value of $E_{\rm b}$. Combining all these tests, we give a conservative
estimate of this latter error of $\pm 30$ meV/O atom, which in turn does not
affect any of the conclusions made in this work.

\section{Stability of sub-surface O}

\subsection{Minimum coverage for O incorporation}

The initial oxidation of the Rh(111) surface proceeds via the dissociative
chemisorption of oxygen in the threefold fcc hollow sites of the basal surface.
\cite{comelli98,loffreda98,pirovano99} O${}_2$ exposure under UHV conditions
leads to two well ordered superstructures, namely a $(2 \times 2)$-O and a
$(2\times 1)$-O phase at coverages of 0.25\,ML and 0.5\,ML, respectively. The
latter coverage was for a long time believed to correspond to a saturated
surface. \cite{comelli98} Yet, after theoretical prediction,
\cite{loffreda98,pirovano99} a $(1\times 1)$-O phase 
($\theta_{\rm tot} = 1.0$\,ML) could be stabilized experimentally
under UHV conditions using atomic O as oxidant, thus demonstrating that the
apparent saturation results from a kinetic energy barrier to O${}_2$
dissociation. \cite{gibson99} This barrier can also be overcome by
enhanced O${}_2$ dosage at moderately elevated temperatures
\cite{peterlinz95,wider99,castner80} or by exposure to more oxidizing carrier
gases like e.g. NO${}_2$. \cite{peterlinz95,gibson99} In all cases, sub-surface
O incorporation or bulk diffusion was reported for higher exposures at
temperatures above  $\sim 400$ K, and as already mentioned a recent experimental
study brought evidence that at least $\sim 0.9$\,ML of oxygen is adsorbed on the
surface, before sub-surface sites become occupied. \cite{wider99}

To address this initial incorporation theoretically we study the stability and
properties of sub-surface oxygen species as function of the total coverage,
$\theta_{\rm tot}$ ($0.25\leq\theta_{\rm tot}\leq 1.0$ ML). We employ 
$(2 \times 2)$ unit-cells and calculate the average adsorption energy 
of fully relaxed structures containing from 
$N_{\rm tot} = 1$ ($\theta_{\rm tot} = 0.25$\,ML) up to 
$N_{\rm tot} = 4$ ($\theta_{\rm tot} = 1$\,ML) oxygen atoms (see Eq. 
(\ref{eq2})). The occupation of sub-surface sites will commence, when a 
structure with one of these O-atoms located below the surface becomes 
energetically more favorable than the most stable one with all oxygens on
the surface at the same total coverage. That is, we compare the stability of 
$(2 \times 2)-((N_{\rm tot}-1) {\rm O}_{\rm on} + {\rm O}_{\rm sub})$/Rh(111)
mixed phases relative to the $(2 \times 2)-(N_{\rm tot} {\rm O}_{\rm on})$/Rh(111)
structure with O${}_{\rm on}$ atoms in fcc hollow sites for 
$N_{\rm tot} = 1,\ldots 4$.

Between the first and second outermost metal layers, there are three different
high-symmetry interstitial sites available for O incorporation. The octahedral
site (henceforth octa) lies just underneath the fcc on-surface site, and one 
tetrahedral site (tetra-I) lies below the hcp on-surface site. A second 
tetrahedral site (tetra-II) is located directly below a first layer metal
atom. Considering both three-fold hollow sites (fcc and hcp) as possible
adsorption sites on the surface, leads then to a many-fold of possible 
structural combinations of how to place the $N_{\rm tot}$
oxygen atoms into the unit-cell, in particular as at most coverages several
symmetry inequivalent possibilities for the same on- and sub-surface site
combination exist.

\begin{figure}
\begin{center}
\psfig{file=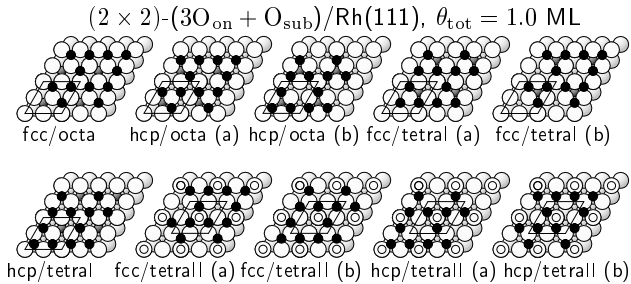,width=1.0\columnwidth,angle=0}
\end{center}
\caption{Top view of all possible on-surface/sub-surface
site combinations at $\theta_{\rm tot} = 1.0$\,ML with one oxygen atom located
below the surface (see text for the explanation of the different sites). Rh =
big spheres (white = surface layer, grey = 2nd layer), O = small spheres (black
= on-surface, grey = sub-surface). Oxygens in tetra-II sites below the first
layer Rh atoms are invisible in this plot and are schematically shown as small
white circles. Symmetry inequivalent occupation of the same kind of on- and
sub-surface sites are denoted with (a) and (b), respectively.}
\label{fig1}
\end{figure}

\begin{figure}
\begin{center}
\psfig{file=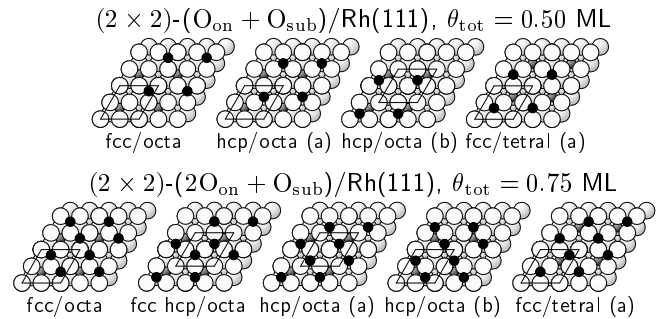,width=1.0\columnwidth,angle=0}
\end{center}
\caption{Top view of selected on-surface/sub-surface
site combinations at $\theta_{\rm tot} = 0.5$\,ML and
$\theta_{\rm tot} = 0.75$\,ML with one oxygen atom located
below the surface (see text). Rh = big spheres
(white = surface layer, grey = 2nd layer), O = small spheres
(black = on-surface, grey = sub-surface).}
\label{fig2}
\end{figure}

As will be shown below, the stability of structures  involving on- and
sub-surface sites becomes only comparable to that of the pure chemisorbed
on-surface phase at $\theta_{\rm tot} = 1.0$\,ML. Thus, only at that coverage
we will address the complete set of possible structural combinations, shown 
in Fig. \ref{fig1}, while at the lower coverages we only exemplify the
stability trends by computing three likely combinations that will become
relevant at a later stage of our discussion. Namely, these are geometries
where on-surface oxygen is in fcc sites and the sub-surface oxygen in either
the octa or the tetra-I sites, fcc/octa and fcc/tetra-I respectively. 
Thirdly, we included the experimentally suggested possibility \cite{wider99}
that oxygen in octahedral sub-surface sites induces a site-switch of the
nearby on-surface oxygens from fcc to hcp, viz hcp/octa. The thus resulting
set of considered structures is shown in Fig. \ref{fig2} for 
$\theta_{\rm tot} = 0.5$\,ML and $\theta_{\rm tot} = 0.75$\,ML. At the
latter coverage we also tested the possibility of a simultaneous occupation
of hcp and fcc on-surface sites and octahedral sub-surface sites, leading to
the fcc hcp/octa structure. Figs. \ref{fig1} and \ref{fig2} also show the
considered symmetry inequivalent structures with the same kind of on- and
sub-surface sites occupation, e.g. hcp/octa (a) and hcp/octa (b).

\begin{table}
\caption{\label{tableI}
Average adsorption energies (in eV/O atom) of Rh(111) geometries
containing O in
on-surface and/or sub-surface sites. The label on/sub indicates the site type
(fcc,hcp for on-surface; octa/tetra-I/tetra-II for sub-surface, see text).
Calculated values for the most stable on-surface adlayer with O occupying fcc
sites (fcc/---) are from Ref. {\protect\onlinecite{pirovano99}}.
For $\theta_{\rm tot} \leq 1.25$\,ML, 0.25\,ML of O is contained
below the surface, for $\theta_{\rm tot} = 2.0$\,ML, 1.0\,ML of
O is contained below the surface.}
\begin{tabular}{c|rccccc}
Sites                           &\multicolumn{6}{c}{$\theta_{\rm tot}$} \\
on/sub                  & 0.25& 0.50 & 0.75 & 1.00 & 1.25 & 2.00 \\ \hline
fcc/---                 & 2.24 & 1.95& 1.66 & 1.40 &             &                \\
---/octa                        &$-1.25$ &               &                &             &                &                \\
---/tetraI              &$-0.78$ &               &                &             &                &                \\
fcc/octa                        &                & 0.36 & 0.75 & 0.81 & 0.69 & 0.56 \\
fcc hcp/octa    &                & 0.68 &               &                &                &             \\
hcp/octa (a)    &                & 0.81 & 1.02 & 0.92 & 0.81 & 0.89 \\
hcp/octa (b)    &                & 0.34 & 0.68 & 0.74 &           &             \\
fcc/tetraI (a) &                 & 1.02 & 1.19 & 1.07 & 0.93 & 1.13 \\
fcc/tetraI (b) &                 &                &             & 0.74 &                  &             \\
hcp/tetraI (a) &                 &                &             & 0.79 &                  &             \\
fcc/tetraII (a)&                 &                &             & 0.91 & 0.77 & 0.83 \\
fcc/tetraII (b)&                 &                &             & 0.91 &                  &             \\
hcp/tetraII (a)&                 &                &             & 0.80 & 0.68 & 0.86 \\
hcp/tetraII (b)&                 &                &             & 0.85 &                  &             \\
\end{tabular}
\end{table}

\begin{figure}
\begin{center}
\psfig{file=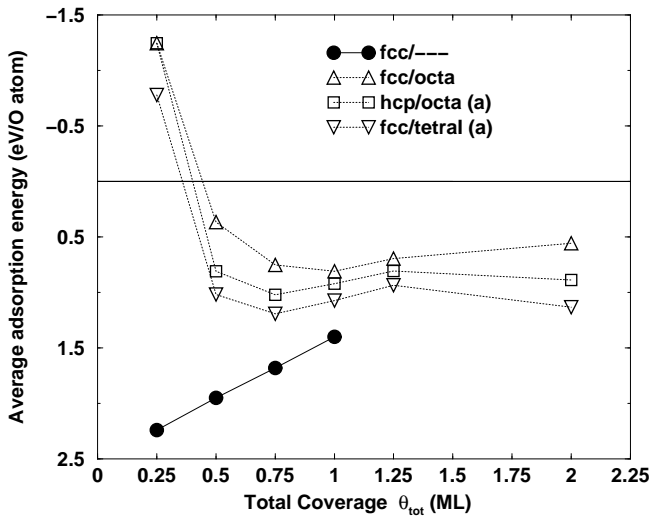,width=0.8\columnwidth,angle=-90}
\end{center}
\caption{Average adsorption energies (in eV/O atom) of
selected Rh(111) geometries containing O in on-surface
and/or sub-surface sites, see text and Table \ref{tableI}.}
\label{fig3}
\end{figure}

The calculated average adsorption energies of all investigated structures
are compiled in Table \ref{tableI}, while Fig. \ref{fig3} additionally
visualizes the trends for three selected on-surface/sub-surface combinations
as described above. Concentrating first on the lowest tested coverage,
$\theta_{\rm tot} = 0.25$\,ML, we find that the occupation of just 
sub-surface sites without any on-surface oxygen is not even exothermic
and by $\approx 3$\,eV/O atom less favorable than adsorption into the
on-surface fcc hollow sites. We have recently shown that this is generally
the case for the closed-packed late $4d$ transition metal surfaces,
which is largely largely due to the significant local expansion of
the metal lattice induced by occupation of sub-surface sites. 
\cite{todorova02} The corresponding cost of distorting the metal lattice
and breaking metal bonds renders sub-surface sites initially always less 
stable than on-surface chemisorption. Yet, upon increasing the
on-surface coverage, repulsive interactions between the
adsorbates drive the adsorption energy down, cf. Fig. \ref{fig3},
until eventually occupation of sub-surface sites might
become more favorable compared to a continued filling
of on-surface sites.

Turning therefore to higher coverages up to 1.0 ML, where we continuously
increase the number of adsorbed on-surface oxygen atoms per unit cell, we
find the average adsorption energies of all mixed structures now to be 
positive, i.e. they should be able to form. Still, in the whole sub-monolayer
regime the values of $E_{\rm ad}(\theta_{\rm tot})$ for the mixed phases 
are significantly lower compared to that of the pure on-surface adsorption
at the same total coverage, reflecting the above described fact, that the
occupation of sub-surface sites is energetically considerably less favorable.
Only at $\theta_{\rm tot} = 1.0$\,ML does the adsorption energy of the 
selected mixed phases approach that of the pure on-surface fcc phase to
within $\approx 0.3$\,eV/O atom (cf. Fig. \ref{fig3}). To make sure that
no other structural on-surface/sub-surface combination would be even slightly
more stable and then eventually more favorable than the pure on-surface 
phase, we tested all of the possible structures shown in Fig. \ref{fig1} at
this particular coverage. As can be seen from Table \ref{tableI} neither of
these configurations leads to a more favorable binding than the pure fcc phase.
Therefore, O incorporation into the Rh(111) surface can only occur at just 
about the completion of the full monolayer coverage on the surface, i.e. 
$\theta_{\rm tot}\approx 1.0$\,ML.

The calculated energy difference of $\approx 0.3$\,eV/O atom between the
pure fcc and the mixed phases for coverages $\theta_{\rm tot} \approx 1.0$\,ML
indicates that sub-surface O penetration into Rh(111) is an energetically 
activated process and that a small, but finite concentration of sub-surface
oxygen can be expected already at on-surface coverages slightly below 1.0\,ML
at elevated temperatures. Wider {\it et al.} have found that extended exposure
of the Rh(111) surface to oxygen at 470 K led to the formation of a sub-surface 
species, though the on-surface coverage remained slightly below 1.0\,ML.
\cite{wider99} In comparison, on Ru(0001) we find the energy difference 
between the pure on-surface and the mixed phases at $\theta_{\rm tot} = 1.0$\,ML
to be $\approx 0.8$\,eV/ O atom, \cite{ruthenium} that is about a factor
of three larger than for Rh(111). Correspondingly, oxygen penetration has at
the prior surface only been reported to occur {\em after} the $(1 \times 1)$-O
phase on the surface has been completed. \cite{stampfl96,boettcher99}

\subsection{Site preferences and site-switch}

Trying to understand the oxygen site preferences in the various
on-surface/sub-surface phases at coverages up to 1.0 ML, we note
first that the low stability of a number of structural combinations
can be understood in terms of electrostatic repulsion between 
the oxygens, i.e. whenever the electronegative oxygens come too
close to each other. This holds for mixed fcc hcp combinations 
(cf. Table \ref{tableI}), as well as for the less stable of the
two symmetry inequivalent possibilities of the same 
on-surface/sub-surface site combination (rendering e.g. the hcp/octa
(b) and fcc/tetra (b) geometries unfavorable, cf. Figs. \ref{fig1}
and \ref{fig2} and Table \ref{tableI}).

Moreover, we find the stability of all other combinations at 
$\theta_{\rm tot}=1.0$ ML to be rather similar, i.e. within 
$\approx 0.2$\,eV/O atom, cf. Table I. This suggests that at
elevated temperatures oxygen incorporation could start initially
in all available sites, and that kinetic factors (like penetration
barriers) more than energetic factors determine which of the 
possible metastable sites get populated first in an experiment
with controlled oxygen dosage. As noted in the introduction, 
a recent analysis of XPD data indicated the presence of a 
sub-surface oxygen species ($\sim 0.1$\,ML) in octahedral sites 
between the first and second metal layer at an almost
oxygen-covered ($\sim 0.9$\,ML) Rh(111) surface after $10^5$
Langmuir O${}_2$ exposure at 470 K. The analysis of the data
yields to the suggested simultaneous occupation of neighboring
on-surface hcp sites, which is different from the otherwise
preferred fcc sites on Rh(111) before oxygen incorporation. 
\cite{wider99}

\begin{figure}
\begin{center}
\psfig{file=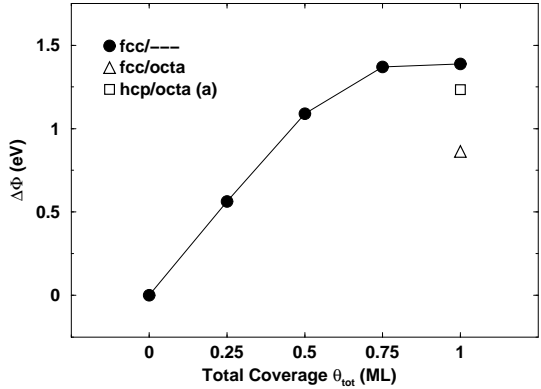,width=0.6\columnwidth,angle=-90}
\end{center}
\caption{Calculated work function change for the 
fcc/octa and hcp/octa (a) structures at $\theta_{\rm tot}= 1.0$\,ML 
(see Fig.~\protect\ref{fig1}). Calculated values for the most
stable adlayers with O occupying only on-surface fcc sites
are from Ref.~\protect\onlinecite{pirovano99}.}
\label{fig4}
\end{figure}

\begin{figure}
\begin{center}
\psfig{file=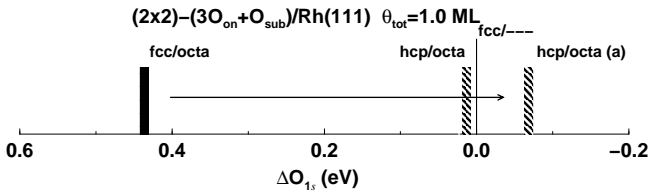,width=0.3\columnwidth,angle=-90}
\end{center}
\caption{
Calculated initial-state on-surface O $1s$ core-level shifts for
the fcc/octa and hcp/octa (a) phases at $\theta_{\rm tot}=1.0$
relative to the O $1s$ level for the adatoms of the O-$(1\times 1)$/Rh(111)
structure at $\theta_{\rm on}=1.0$ ML, with O occupying on-surface
fcc sites (fcc/---). The label on/sub indicates the on- and
sub-surface adsorption sites, respectively. There are two symmetry
inequivalent O adatoms in the hcp/octa phase, which is why two levels
are shown for this structure.}
\label{fig4b}
\end{figure}

This intriguing finding is in line with our calculations, which
indeed indicate, that {\em if} sub-surface oxygen was to occupy
octahedral sites, then the site-switch combination hcp/octa
would be considerably more favorable compared to the fcc/octa
combination with the on-surface oxygens in their normal adsorption
sites, cf. Fig. \ref{fig3}. 

In seeking for a qualitative explanation for the stability of the
hcp  on-surface sites upon occupation of octahedral sites,
and consistent with the discussion on the preferred fcc site 
adsorption begun in earlier papers, \cite{pirovano99,pirovano01}
we examine the calculated changes in the work-function, $\Delta\Phi$, 
as well as the difference in O $1s$ binding energy of on-surface
O-atoms, $\Delta{\rm O}_{1s}$, of the mixed fcc/octa and hcp/octa
phases at $\theta_{\rm tot}=1.0$ ML relative to the values for the 
stable $(1\times 1)$-O/Rh(111) structure, with O just occupying fcc
sites. The calculated $\Delta\Phi$ and $\Delta{\rm O}_{1s}$ values
are shown in Figs. \ref{fig4} and \ref{fig4b}, respectively. Initially,
the work function rises as a function of the coverage, while O remains
in the surface fcc sites; at $\theta_{\rm tot}\approx 0.75$ ML it
reaches a saturation value. Such an increase in the work-function
upon O chemisorption reflects the high electronegativity of the adspecies
that results in an induced inward dipole moment, i.e., with the negative
charge at the vacuum side of the surface. The saturation at higher 
coverages is then a consequence of the dipole-dipole interaction giving
rise to a depolarization with decreasing O-O distance. 

The initial incorporation of oxygen in the fcc/octa and hcp/octa phases
leads to a decrease in the work function compared to the saturation value
at 1.0 ML with O just occupying on-surface fcc sites. This decrease is 
considerably smaller for the hcp/octa structure, cf. Fig. \ref{fig4};
where it should also be noticed that the average bond length, 
O$_{\rm on}$-Rh$_{1}$ and interlayer spacing, $\bar d_{\rm O_{\rm on}-Rh_1}$, 
between chemisorbed O and Rh atoms at the surface are practically 
independent of the occupation of the sites at a given coverage, cf.
Tables  \ref{tableII} and \ref{tableIII}. Moreover, the calculated 
initial-state O $1s$ shifts schematically shown in Fig. \ref{fig4b}, 
indicate that the O $1s$ levels of adatoms in the hcp/octa geometry 
are $\sim 0.5$ eV {\it less} bound compared to corresponding levels
of oxygen atoms in the fcc/octa geometry. This reflects a larger 
electrostatic repulsion at the hcp sites, where a somewhat more
negatively charged O would be adsorbed, which in turn correlates with
the smaller decrease in the work function for the preferred site-switch
phase.

On comparing the values of $\Delta\Phi$ and $\Delta{\rm O}_{1s}$ for the
hcp/octa and fcc/octa phases with those for the pure fcc adsorption in 
Figs. \ref{fig4} and \ref{fig4b} it can thus be suggested that a similarly 
charged O would sit in hcp sites (rather than in fcc sites) upon 
octahedral sub-surface occupation compared to the oxygens adsorbed in fcc
sites before O penetration. Hence, similarly to the pure on-surface fcc 
adsorption, \cite{pirovano99,pirovano01} we suggest that a stronger 
{\it ionic} bonding favors the hcp sites in the site-switch phases at the 
onset of O penetration ($\theta_{\rm tot}\approx 1.0$ ML).

Although we can thus rationalize the higher stability of the hcp/octa 
site-switch phase suggested by Wider {\em et al.} \cite{wider99} compared
to the fcc/octa phase, we nevertheless stress that the occupation of
octahedral sub-surface sites is according to our calculations only of a
metastable character, as the fcc/tetra-I structural combination
is consistently energetically more favorable over the complete
tested coverage range, cf. Fig. \ref{fig3}.

\subsection{Continued oxidation: island formation and trilayer shift}

We investigate the continued oxidation after sub-surface O
has been initially incorporated in the coverage region
$1.25\leq\theta_{\rm tot}\leq 2.0$ ML by calculating the
average adsorption energy at both 1.25\,ML ($\theta_{\rm sub}=0.25$ ML)
and 2.0\,ML ($\theta_{\rm sub}=1.0$ ML) coverages. Again,
we find that the site combinations hcp/octa and fcc/tetra-I
give the highest adsorption energies, while the occupation
of tetra-II sites leads to energetically slightly less favorable
phases, cf. Table I. In the following we will restrict our discussion 
to the two most stable phases, i.e. hcp/octa and fcc/tetra-I.

\begin{table}[ht]
\caption{\label{tableII}
Calculated structural parameters in {\AA} for the hcp/octa and
fcc/tetra-I phases at $\theta_{\rm tot}=1.25$ ML. O atoms and all
Rh atoms in the two outermost layers were allowed to relax. For
the interlayer distances, the center of mass of each layer is
used. Numbers in parenthesis correspond to bulk values, which
were fixed. O$_{\rm on,sub}$-Rh$_{1,2}$ indicate (averaged) bond
lengths to first and second layer Rh atoms, and $\Delta z_{{\rm Rh}_1,2}$
and $\Delta z_{{\rm O}_{\rm on}}$ the magnitude of the buckling
of the outermost Rh layers and of the on-surface O adlayer,
respectively. The lateral displacements, radially away from the
ideal lattice positions, are denoted by $\Delta r_{{\rm Rh}_{1,2}}$,
and $\Delta r_{{\rm O}_{\rm on}}$, respectively.}
\begin{tabular}{lccc}
                                                                                  & hcp/octa & fcc/tetra-I\\ \hline
O$_{\rm on}$-Rh$_{1}$                     &  1.94        &       1.95             \\
O$_{\rm sub}$-Rh$_{1}$                    &  2.15        &       2.00             \\
O$_{\rm sub}$-Rh$_{2}$                    &  2.21        &       1.94             \\
$\bar d_{\rm O_{\rm on}-Rh_1}$  &  1.14  &       1.15             \\
$\bar d_{12}$                                             &  2.79        &       2.78             \\
$\bar d_{23}$                                             &  2.30        &       2.32             \\
$\bar d_{34}$                                             & (2.213)      & (2.213)        \\
$\Delta z_{{\rm Rh}_1}$                   &  0.30        &       0.23             \\
$\Delta z_{{\rm Rh}_2}$                   &  0.26        &       0.30             \\
$\Delta z_{{\rm O}_{\rm on}}$     &  0.18 &      0.18             \\
$\Delta r_{{\rm Rh}_1}$                   &  0.06        &       0.08             \\
$\Delta r_{{\rm Rh}_2}$                   &  0.03        &       0.00             \\
$\Delta r_{{\rm O}_{\rm on}}$     &  0.07        &       0.07             \\
\end{tabular}
\end{table}

\begin{table}[ht]
\caption{\label{tableIII}
Calculated structural parameters in {\AA} for the hcp/octa,
fcc/tetra-I, and fcc/tetra-I$_{\rm fault}$ phases at
$\theta_{\rm tot}=2.0$ ML. O atoms and all Rh atoms in the
two outermost layers were allowed to relax. Numbers in
parenthesis correspond to bulk values, which were fixed.
O$_{\rm on,sub}$-Rh$_{1,2}$ indicate bond lengths to first
and second layer Rh atoms, and $\bar d_{ij}$ interlayer
spacings.}
\begin{tabular}{lccc}
                        & hcp/octa & fcc/tetra-I & fcc/tetraI$_{\rm fault}$\\\hline
O$_{\rm on}$-Rh$_{1}$                   &       1.96      &      1.97            &                      1.97                      \\
O$_{\rm sub}$-Rh$_{1}$                  &       2.10      &      2.05            &                      2.06                      \\
O$_{\rm sub}$-Rh$_{2}$                  &       2.51      &      2.01            &                      2.02                      \\
$\bar d_{\rm O_{\rm on}-Rh_1}$& 1.19      &      1.19            &                      1.19                      \\
$\bar d_{12}$                                           &       3.36      &      3.34            &                      3.35                      \\
$\bar d_{23}$                                           &       2.14      &      2.17            &                      2.18                      \\
$\bar d_{34}$                                           & (2.213)  &    (2.213)  &                (2.213)                 \\
\end{tabular}
\end{table}

Tables \ref{tableII} ($\theta_{\rm tot} = 1.25$\,ML) and
\ref{tableIII} ($\theta_{\rm tot} = 2.0$\,ML) show that
the geometrical changes in response to sub-surface penetration
of oxygen are significant and it is therefore not accurate to
assume that the Rh(111) lattice will remain essentially
undisturbed upon the occupation of sub-surface sites.
For both hcp/octa and fcc/tetra-I phases at 1.25 ML (0.25 ML
below the surface), the mean outermost substrate interlayer
spacing, $\bar d_{12}$, expands by about $\sim 22\%$, and
$\sim 26\%$, relative to the corresponding value for the
bulk-terminated surface, respectively (cf. Table \ref{tableII}).
Prior to the present work, we have calculated an increase of
$\bar d_{12}$ from 2.21\,{\AA} for the clean surface to 2.37{\AA}
for the $(1\times 1)$-O/Rh(111) structure with oxygens in fcc
sites, \cite{pirovano99} which means that an additional
$\sim 15-19\%$ expansion is induced by the 0.25 ML of sub-surface
oxygen.

\begin{figure}
\begin{center}
\psfig{file=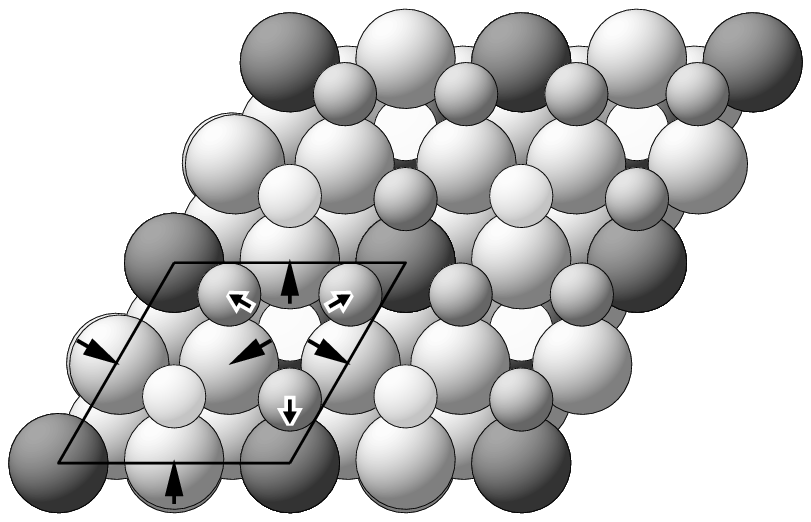,width=1.0\columnwidth,angle=0}
\psfig{file=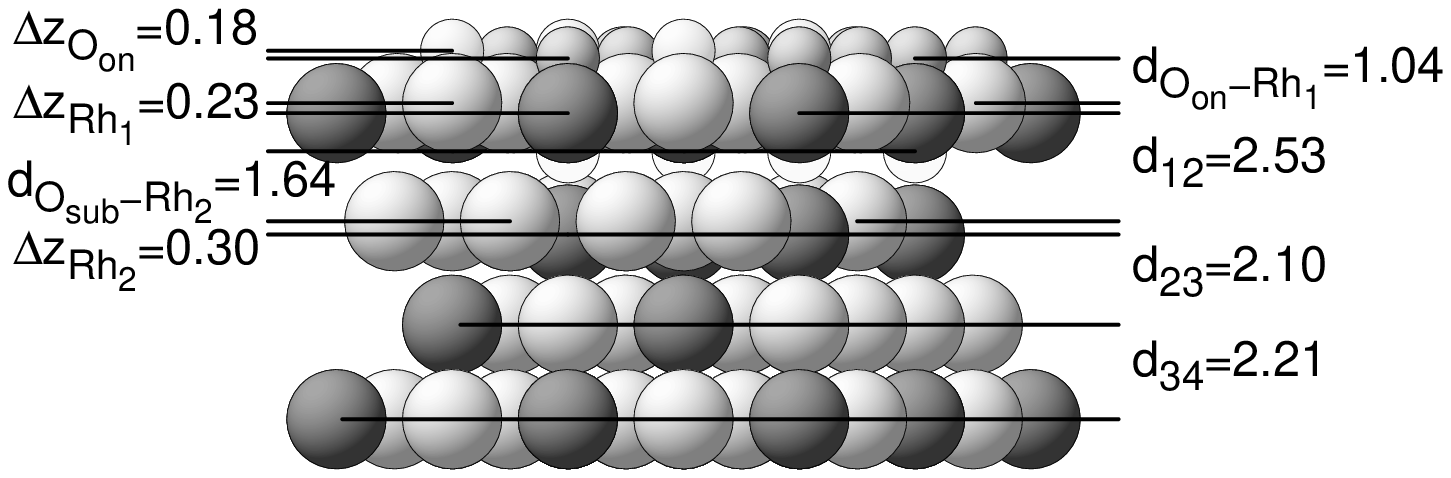,width=1.0\columnwidth,angle=0}
\end{center}
\caption{Top and side view of the fcc/tetra-I geometry at
$\theta_{\rm tot} = 1.25$\,ML. Small and large spheres
represent oxygen and Rh atoms respectively, where those
lying in the same plane and equivalent under the threefold
rotation symmetry have the same color. The arrows indicate
the direction of reference for the atomic in-plane displacements
$\Delta r_{{\rm Rh}_{1,2}}$, and $\Delta r_{{\rm O}_{\rm on}}$
of Table~\protect\ref{tableII}. Distances are in {\AA}.}
\label{fig5}
\end{figure}

For the $(2 \times 2)$-($4\rm O_{\rm on}+\rm O_{\rm sub}$)/Rh(111)
phases at 1.25\,ML, symmetry allows for (in part considerable)
buckling of the outermost Rh and on-surface O layers, i.e. quite
distinct local contractions and expansions together with lateral
shifts (cf. Table \ref{tableII}), which can most often be understood
as a local expansion of the metal lattice around the occupied
sub-surface site. \cite{todorova02} In view of keeping the paper
within a limited length, we show only one example. For instance,
in the fcc/tetra-I geometry, cf. Fig. \ref{fig5}, the first layer
Rh atoms above the occupied sub-surface site, are pulled out of the
surface by $\sim 0.23$\,{\AA}, while the neighboring Rh atoms in
the second substrate layer are pushed inwards by $\sim 0.30$\,{\AA}.
Also, there are local vertical displacements of the O adatoms
($\sim 0.18$\,{\AA}), and considerable lateral shifts radially away
from the ideal lattice positions for both substrate and adsorbate
atoms.

\begin{figure}
\begin{center}
\psfig{file=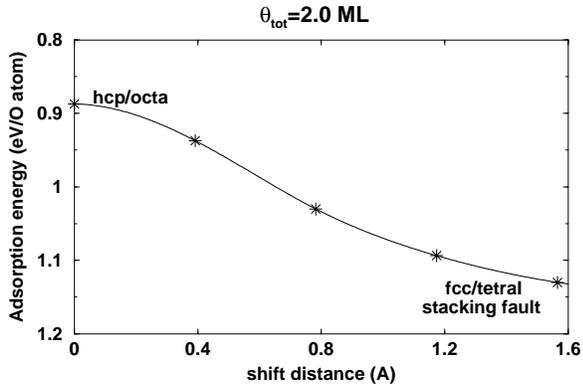,width=0.6\columnwidth,angle=-90}
\end{center}
\caption{Calculated average adsorption energy during the
registry shift of the O${}_{\rm on}$-Rh-O${}_{\rm sub}$
trilayer along the $[\bar 2 1 1]$ direction over the Rh(111)
substrate. The initial hcp/octa phase ends in a fcc/tetra-I
configuration with a fault in the stacking sequence of the
surface Rh layer after shifting the trilayer by
${\frac{a_0}{3}}\sqrt{3/2}$ ($a_0=3.83$ {\AA}).}
\label{fig7}
\end{figure}

\begin{figure}
\begin{center}
\psfig{file=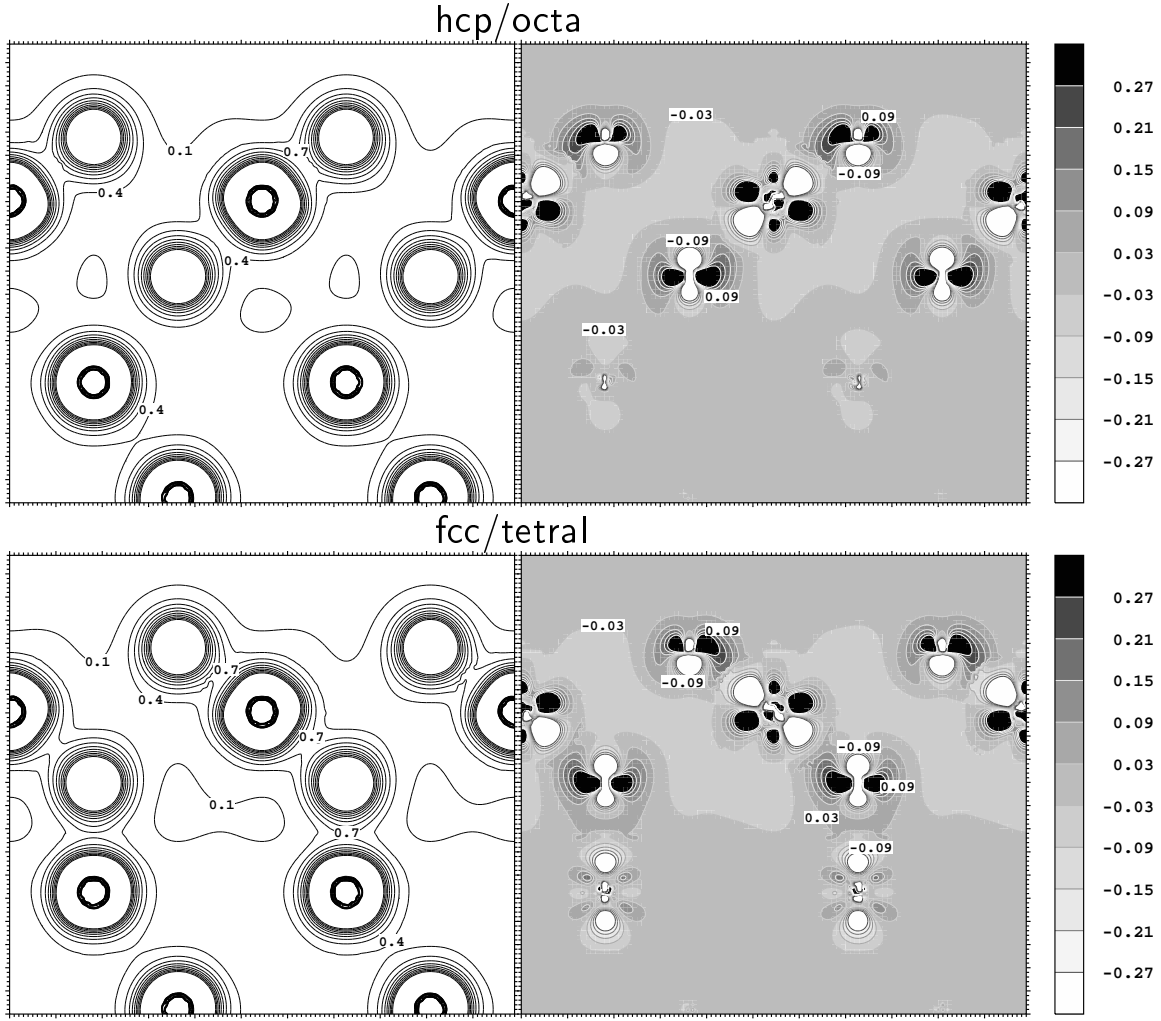,width=0.8\columnwidth,angle=90}
\end{center}
\caption{Left: Contour plots of constant electron density in a
$[\bar 2 1 1]$ plane perpendicular to the (111) surface of
$(2\times 2)-(4 \rm O_{\rm on}+ 4 \rm O_{\rm sub}$)/Rh(111) structures
with oxygens in hcp/octa (top) and fcc/tetra-I (bottom) sites.
Right: Plot of the difference charge density of the same
plane, i.e. where the charge density of the clean Rh(111)
surface (with distances between Rh atoms as in the chemisorbed system)
and those of the isolated oxygens have been subtracted from the
electron density shown to the left. The distance between contours
is 0.3 $e/${\AA}${}^3$ (left) and 0.06 $e/${\AA}${}^3$ (right). In
both cases, the improved bonding of the trilayer to the
underlying Rh(111) substrate can be seen as an increase in
the bonding charge density to the second layer Rh atoms.}
\label{fig6}
\end{figure}

The expansion of the first Rh layer distance becomes with $\approx 51$\%
even more pronounced at $\theta_{\rm tot} = 2.0$\,ML (1\,ML O below the
surface), cf. Table \ref{tableIII}. Despite this significant distortion
of the metal lattice, we find increased adsorption energies for both
the hcp/octa and fcc/tetra-I phases compared to the low sub-surface O
coverage at $\theta_{\rm tot} = 1.25$\,ML, cf. Fig. \ref{fig3}. This
reflects an attractive interaction between the sub-surface species, which
implies that the latter have a tendency to form {\em sub-surface islands}
(i.e. nucleate) with a local $(1 \times 1)$ periodicity. For both the hcp/octa and
fcc/tetra-I phase, the structure of these islands can be viewed as an
O${}_{\rm on}$-Rh-O${}_{\rm sub}$ trilayer on top of a Rh(111) substrate,
cf. Fig. \ref{fig6}. Interestingly, the internal geometry of the strongly
bound trilayer is almost identical for both phases, cf. Table \ref{tableIII}
and Fig. \ref{fig6}; it is only the coordination to the underlying
substrate which is different in both cases. In fact, we even find that the
higher energy hcp/octa configuration is unstable against a registry shift of
this whole trilayer along the $[\bar 2 1 1]$ direction. The calculated
average adsorption energy along this {\it sliding} of the trilayer over
the Rh(111) surface is shown in Fig. \ref{fig7}. We have fully optimized
the structures at the calculated points along the {\em barrierless}
displacement, starting from the hcp/octa structure. At the end of the shift
by $\frac{a_0}{3}{\sqrt{3/2}}$ ($a_0=3.83$ {\AA}, Rh lattice constant
\cite{pirovano99}), the on-surface oxygen atoms are located in fcc and the
sub-surface oxygen atoms in tetra-I sites, but the surface Rh layer inside
the trilayer does not continue the fcc lattice stacking, but is now located
in a stacking fault position. The calculated structural parameters
(interlayer spacings and bond lengths) between these two fcc/tetra-I$_{\rm fault}$
and fcc/tetra-I geometries have remained virtually unchanged (cf. Table
\ref{tableIII}), and the average binding energies differ by only $\sim 3$
meV/O atom, i.e. they are degenerate within our calculational uncertainty,
which is plausible as the two structures differ in fact only by a 60${}^o$
rotation of the trilayer with respect to the underlying substrate. We
have similarly compared the geometries and average binding energies of both,
fcc/tetraI and fcc/tetraI$_{\rm fault}$ structures at 1.25 ML and also found no
significant difference; we find a difference of $\sim 1$ meV/atom in the
calculated binding energies. 

The reason behind the $\sim 0.2$\,eV/O atom ($\sim 0.4$\,eV/unit cell)
preference for the fcc/tetra-I
structure compared to the hcp/octa phase can be found when analyzing
the coupling of the formed trilayer to the underlying Rh(111) substrate.
Whereas we consistently find O-Rh bondlengths of $\sim 2.0$\,{\AA}
for both on- and sub-surface oxygens in the majority of tested structures,
the O${}_{\rm sub}$-Rh${}_2$ bondlength of the sub-surface oxygen to
the second layer Rh atom in the hcp/octa phase at $\theta_{\rm tot} = 2.0$\,ML
is significantly larger (2.51\,{\AA}). This points to a weak coupling
of the trilayer to the Rh(111) substrate in this geometry, which can also
be seen in the density plots shown in Fig. \ref{fig6}, where
the incorporation of sub-surface O in octahedral sites is found
to induce only very small changes on the valence charge of the second
layer Rh atoms. These changes are considerably larger for the fcc/tetra-I
configuration, suggesting that the energetic preference of this structure
is primarily due to an improved coupling of the formed trilayer to the
underlying substrate.

Hence, even if sub-surface oxygen was initially incorporated into the
metastable octahedral sites as suggested by experiment, \cite{wider99}
it would due to this instability transform into the fcc/tetra-I
configuration upon continued oxygen penetration into the Rh(111) surface. 
It is interesting to notice that on the Ru(0001) surface \cite{reuter02a,reuter02b}
a plausible oxidation pathway has been identified, in which after the 
formation of such trilayers a phase transformation into the rutile
RuO${}_2$(110) structure was obtained for local oxygen coverages exceeding
5\,ML. The local formation of such precursing structures via the
agglomeration of sub-surface oxygen atoms below the surface that follow
the initial oxygen incorporation, could therefore be a more general 
phenomenon in the oxidation of transition metal surfaces. Once the local
oxygen coverage exceeds a critical value, these precursor configurations 
will then undergo a structural change and actuate the formation of bulk
oxide phases. In this respect we notice that already in the preferred 
fcc/tetra-I configuration the Rh first layer atoms are sixfold coordinated
to oxygens and the oxygen atoms in the tetrahedral site fourfold coordinated 
to metal atoms, i.e. they exhibit identical local coordinations as in the
most stable, corundum-structured Rh${}_2$O${}_3$ bulk oxide.

\section{Summary}

In conclusion we have presented a density-functional theory
study addressing the initial penetration of oxygen into the
Rh(111) surface. Due to the large local expansion of the metal
lattice induced by the occupation of sub-surface sites, O
chemisorption on the surface is initially significantly more
favorable. In agreement with recent experimental findings,
we therefore find oxygen penetration to occur only after
the adsorption of {\it almost} a full monolayer on the surface.
A particularly interesting result is that the calculated
energy difference of $\sim 0.3$ eV/atom between the pure 
on-surface and the mixed on/sub-surface phases at 
$\theta_{\rm tot} = 1.0$\,ML is about a factor of three 
smaller than that for Ru(0001). Oxygen penetration has at the
Ru surface only been reported to occur {\em after} the 
$(1 \times 1)$-O phase on the surface has been completed.

The experimentally suggested occupation of octahedral sites
in connection with a site-switch of the on-surface oxygens
from fcc to hcp sites is indeed found to be initially possible
as a metastable configuration. Increased sub-surface O
incorporation will then lead to the agglomeration of 
sub-surface species. The hcp/octa phase is unstable against a 
registry shift of the O${}_{\rm on}$-Rh-O${}_{\rm sub}$
outermost layers by which the initial hcp/octa phase ends
in a fcc/tetra-I configuration with a fault in the stacking 
sequence of the surface Rh layer. In this most stable phase, 
the on- and sub-surface oxygens occupy fcc and tetra-I sites,
respectively, and metal atoms are sixfold coordinated to oxygens
while the sub-surface oxygen atoms in tetrahedral sites are 
fourfold coordinated to metal atoms.

The similarities between these results and those of a simultaneous
theoretical study of the oxidation of the Ru(0001) surface provide 
a basis for the interpretation of the role of oxygen incorporation,
and nucleation as precursors of the final bulk oxide structure. 
Addressing the phase transformation to the Rh${}_2$O${}_3$ bulk 
oxide will thus be of considerable future interest, completing the 
atomistic pathway of oxide formation at the Rh(111) surface.

\section{Acknowledgements}

We thank T. Greber, J. Osterwalder, A. Seitsonen, C. Stampfl, 
F. Illas, and  M.E. Grillo for helpful discussions.

\onecolumn

\twocolumn


\begin{references}
\bibitem{ruthenium}
The value of $\approx 0.8$ eV/O atom was calculated by comparing the
average adsorption energy of
$(2\times2)$-($3{\rm O}_{\rm on}+{\rm O}_{\rm sub}$)/Ru(0001)
ordered structures with that of a full adlayer on the surface
[$(1\times1)$-O/Ru(0001), with O on hcp sites]. The coadsorbate
structures considered on Ru(0001) are analogous to the fcc/tetraI
(a)-(b), and the hcp/oct (a)-(b) of Fig. {\protect\ref{fig1}}.
Calculations were performed within the same computational scheme
as employed in the present study, modelling the metal surface by a
six layer slab. Details of the used FP-LAPW basis set are those
described in Refs. {\protect\onlinecite{reuter02b,lizzit01}}.
\end{references}
\end{document}